\documentclass{midl} 
\usepackage{multirow}
\usepackage{booktabs}
\usepackage{colortbl}
\usepackage{hhline}
\usepackage{tikz}
\usetikzlibrary{shapes,arrows}
\usetikzlibrary{intersections}
\usepackage{svg}
\newcommand{\Mtb}{\textit{Mtb.}}
\newcommand*{\Scale}[2][4]{\scalebox{#1}{$#2$}}%
\usepackage{amsmath}

\DeclareMathOperator*{\mbfx}{\mathbf{x}}
\DeclareMathOperator*{\mbfy}{\mathbf{y}}
\DeclareMathOperator*{\mbfz}{\mathbf{z}}

\usepackage{mwe} 
\jmlrvolume{-- Under Review}

\title[Lung Imaging \& Disentangled Representations]{Translational Lung Imaging Analysis Through Disentangled Representations}

\midlauthor{\Name{Pedro M. Gordaliza
\nametag{$^{1}$}} \Email{pmacias@ing.uc3m.es}\\
\Name{Juan José Vaquero
\nametag{$^{1}$}} \Email{jjvaquer@uc3m.es}\\
\Name{Arrate Muñoz-Barrutia\nametag{$^{1}$}} \Email{mamunozb@ing.uc3m.es }\\
\addr $^{1}$ Depart. de Bioingeniería e Ing. Aeroespacial, Universidad Carlos III de Madrid, Leganés, Spain\\
}

\begin{document}

\maketitle

\begin{abstract}
The development of new treatments often requires clinical trials with translational animal models using (pre)-clinical imaging to characterize inter-species pathological processes. Deep Learning (DL) models are commonly used to automate retrieving relevant information from the images. Nevertheless, they typically suffer from low generability and explainability as a product of their entangled design, resulting in a specific DL model per animal model. Consequently, it is not possible to take advantage of the high capacity of DL to discover statistical relationships from inter-species images. \\
To alleviate this problem, in this work, we present a model capable of extracting disentangled information from images of different animal models and the mechanisms that generate the images. Our method is located at the intersection between deep generative models, disentanglement and causal representation learning. It is optimized from images of pathological lung infected by Tuberculosis and is able: a) from an input slice, infer its position in a volume, the animal model to which it belongs, the damage present and even more, generate a mask covering the whole lung (similar overlap measures to the \textit{nnU-Net}), b) generate realistic lung images by setting the above variables and c) generate counterfactual images, namely, healthy versions of a damaged input slice.   
\end{abstract}

\begin{keywords}
Representation learning, disentanglement, translational models, lung, CT.
\end{keywords}

\section{Introduction}
The longitudinal characterization of animal models is crucial during (pre-)clinical drug trials. To characterize disease progression meaningfully, we need to have the capacity to extract comparable biomarkers in similar phases of the disease progression. Besides, we need to prove the existence of similar pathophysiological mechanisms modulating common causal factors that give rise to the variability of trial outcomes. \\
In this context, medical imaging techniques enable the extraction of indicators (imaging biomarkers) from \textit{in vivo} studies \cite{Willmann2008MolecularDevelopment}. For example, the number of \textit{Mycobacterium tuberculosis} (\Mtb) colonies present in a subject can be inferred from the damaged lung volume in an image of a human, primate, or mouse \cite{Yang2021OneChemotherapeutics}.\\
The images contain meaningful information to interpret the mentioned physiological process. However, their manual analysis is tedious, and automation is advantageous to process the vast amount of data produced during the trials. Thus, developing Artificial Intelligence (AI) systems that can not only automate the extraction of particular markers for each animal model (e.g., the damaged lung volume) but are also capable of inferring the common agents of such particular indicators (e.g., bacterial burden) is essential.

Although AI, especially Deep Learning (DL), has eased the process \cite{Zhou2021APromises, Hinton2018DeepCare}, some design premises has lessened its inference capabilities. In particular, DL models excel at extracting the statistical dependence between input-output pairs, i.e.,$(x_i, y_i) \in \mathcal{X,Y}$, from assumed \textit{independent and identically distributed (i.i.d.)} observational data \cite{Peters2017ElementsAlgorithms}.

Such success has leaned the model designs towards an insufficient representation learning strategy \cite{Bengio2013RepresentationPerspectives}. Namely, discovering statistical dependence between specific data pair samples is prioritized rather than understanding the physical model generating the whole data population (e.g., physiological mechanisms).

Since the i.i.d. assumption is fragile, 
well-known distribution shifts \cite{Castro2020CausalityImaging} between data employed at training, validation and test phases, and \textit{real-world} data are usual. Under this scenario, the models tend to learn correlated representations that only hold for specific environments or domains, namely \textit{spurious correlations} \cite{Arjovsky2020InvariantMinimization}. 
Since (as a mantra) \textit{correlation does not imply causation}, such flaws cause ruinous effects \cite{DeGrave2021AISignal, Roberts2021CommonScans} for generalisation, transferability and explainability purposes \cite{Scholkopf2021TowardLearning}.

More formally, naive DL models maximize a joint distribution, $p(X, Y)$ or $p(X)$ (self-supervision), characterized by an entangled representation of the input. Namely, if $X$ and $Y$ correlate during training without necessarily deriving from a causal representation ($X \rightarrow Y$), $p(X, Y)$ can adopt numerous factorization forms that are domain-specific \cite{Goyal2021InductiveCognition}. Thus, forcing to implement independent models even for related domains (in our case, lung CT images of TB animal models). Such models are put in common through \textit{posthoc} analysis, losing possible data synergies.

In general, learning strategies mitigate this issue by shrinking the $p(X, Y)$ solutions space. To this aim, models are enriched injecting inductive biases (e.g., CNNs assume spatial correlation \cite{Dumoulin2018ALearning}),
to facilitate the discovery of more meaningful and disentangled representations \cite{Liu2021ADomain}. These strategies resemble human cognition. Since, humans arrange the proper biases to extract a limited number of relevant factors transferable among different environments \cite{Pearl2011Causality:Edition}.

AI systems design can follow a similar causal perspective. Namely, specific biases can be introduced to shrink the solution space. Thus, in this work, we consider the bias: a) the strongly hierarchical nature of the human visual system and b) the data generation process. Such an approach intends to mimic the radiologists' tasks, who take into account specific patient factors (i.e., clinical history, sex, age) beyond the image \textit{per se}.
This approach yields more effective disentangled representations of the input \cite{Scholkopf2021TowardLearning}.

In particular, we intend to identify the unique mechanisms that govern the generation of translational imaging of lung Computed Tomography (CT) images and their corresponding segmentation masks (\figureref{fig:DAG}). We employ three different animal models (mouse, primate and human) infected by  \Mtb~\cite{Pai2016Tuberculosis}. From a simplified radiological point of view,  mammals' lungs share texture and shape features. We model these shared attributes as an effect of the same causative factors, e.g., bacterial load (see Appendix \ref{ap:cycle}).\\
To prove the benefits of our strategy, we show how after optimizing the model employing a small limited number of volumes, our design can: 
\begin{itemize}
\vspace{0em}
    \item Produce a very accurate reconstruction of the input images and generate  suitable segmentation masks (\figureref{fig:lungs_seg}, \tableref{tbl:optimiza_results}). 
    \vspace{-0.5em}
    \item Generate new realistic images of the three different animal models controlling the lung damage on each, which implies the proper characterization of the disentangled variables (\figureref{fig:generated}).
     \vspace{-0.5em}
    \item Generate counterfactual images of damaged lungs \cite{Schutte2021UsingImages, Cohen2021GifsplanationX-rays}. Namely, the model can capture the meaningful representations of an input image and convert it into a healthy version by intervening on the damage variable.  
\end{itemize}

\section{Methods}\label{sec:Methods}

\begin{figure}[htbp]
\floatconts
  {fig:subfigex}
  {\caption{\small (a) Direct Acyclic Graph (DAG) representing the generation of pathological lung CT images $\mbfx$, and their segmentation masks $\mbfy$. Both generated from a latent variables hierarchy at different resolutions scales, $K$, governed by three factors, i.e., animal model, $A$, the relative position of the axial slice, $S$, and the estimated lung damage caused by \Mtb, $D$. (b) Summarized architecture: Blue and pink represent the image and mask generation branches, $\bigoplus$ features concatenation and $\bigotimes$ $p_\theta$, $q_\phi$ parameters combination in training. The encoder is not present for image generation (Section \ref{ssec:genration}). Counterfactual images arise inferring and setting some values at the deeper representation level $\mathbf{z}^0$ (Section \ref{ssec:counter}).}}
  {%
    \subfigure[\small Direct Acyclic Graph (DAG)]{
    \label{fig:DAG}%
      \includegraphics[width=0.33\linewidth, height=0.135\textheight]{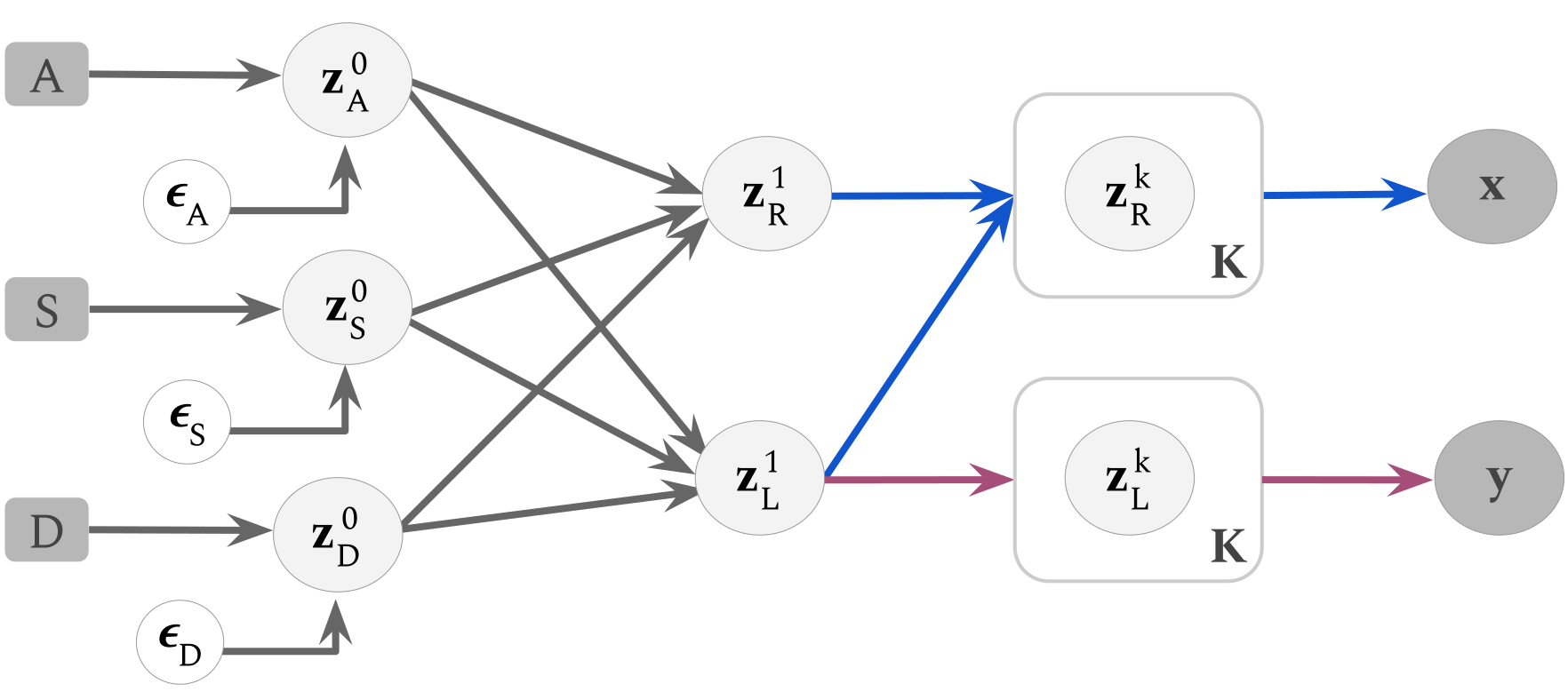}}%
    \quad
    \subfigure[\small Summarized Architecture]{\label{fig:arq}%
      \includegraphics[width=0.5\linewidth, height=0.2\textheight]{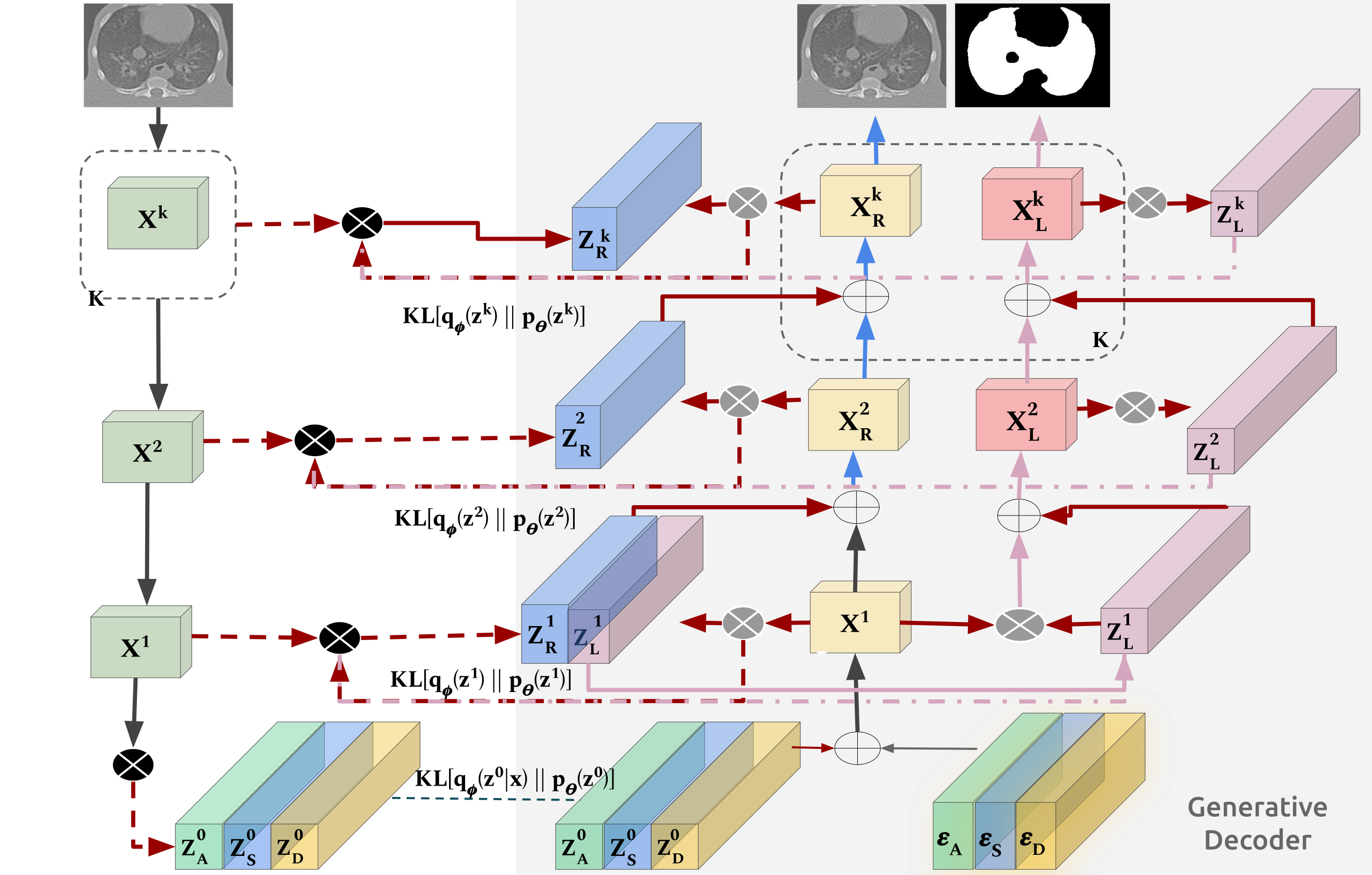}}
  }
\end{figure}

We define a generative model in which the high dimensional texture and shape features that can be extracted from lung CT images and their corresponding segmentation masks are a result of the causal Direct Acyclic Graph (DAG) presented in \figureref{fig:DAG}.
\vspace{0em}
\vspace{-1em}

The proposed DAG simplify the physical image generation for obvious reasons. All the possible elementary causative factors (e.g., specific scanner, comorbidities, subject age, sex) are reduced to three: the animal model, $A$,  the observed lung axial slice, $S$, and the lung damage, $D$. The causative factors are modelled as three groups of independent variables, $\mathbf{z}^0$, under the noise term, $\epsilon_{\{A,S,D\}}$, which comprises noise and unconsidered variables. The primary variables govern the generative process, which follows a part-whole hierarchy \cite{Hinton2021HowNetwork} from low-level representations of the texture and shape features, $\mathbf{z}^1$, to high dimensional ones, $\mathbf{z}^k$, the observed image, $\mbfx$ and the segmentation mask, $\mbfy$. This part-whole hierarchy resembles brain columns functioning \cite{Locatello2020Object-CentricAttention, Devlin2018BERT:Understanding}. Variable superscripts, $\mathbf{z}^k$, symbolize hierarchy levels at the DAG.\\
The plate notation at the DAG represents such upsampling generation. The DAG implements two paths diverging at the first hierarchy level (shared representation path), $\mathbf{z}^1$. The division forces, during optimization, to generate a disentangled representation of shape, $\mathbf{z}_L$ and texture, $\mathbf{z}_R$. CT images depend on shape and texture variables (blue path), while the segmentation masks only depend on shape variables (pink path). Then, assuming the independence of the noise terms,  the \textit{independent causal mechanism} (ICM) principle is fulfilled \cite{Scholkopf2021TowardLearning} and the following disentangled factorization arise:
\begin{equation}\label{eq:factorization}
    p(\mbfx,\mbfy, \mbfz) = p(\mbfx|\mathbf{z}^K_R)p(\mbfy|\mathbf{z}^K_L)p(\mathbf{z}_R^k, )p(\mathbf{z}_L^k)p(\mathbf{z}_R^2|\mathbf{z}_R^1,\mathbf{z}_L^1)p(\mathbf{z}_{R}^1, \mathbf{z}_{L}^1|\mathbf{z}^0)p(\mathbf{z}^0),
\end{equation}

\begin{align}\label{eq:latent_scales}
p(\mathbf{z}_R^k)=\prod_{k=3}^{K}p(\mathbf{z}^k_R|\mathbf{z}^{k-1}_R); \quad  \qquad        &  p(\mathbf{z}_L^k)=\prod_{k=2}^{K}p(\mathbf{z}^k_L|\mathbf{z}^{k-1}_L); \quad   p(\mathbf{z^0}) = p(\mathbf{z}_A^0)p(\mathbf{z}_S^0)p(\mathbf{z}_D^0)          
\end{align}

\subsection{Model optimization}

For the above equations, each conditional distribution is parametrized by depthwise convolutional decoders. The parameters $\theta$, leverages a high capacity model (\figureref{fig:arq} allowing to characterize the unobservable causes of variation ($\mathbf{\epsilon}$) consistent with the available data (in our case, lung CT images) \cite{Peters2017ElementsAlgorithms, Pawlowski2020DeepInference}. Once the model is optimized, it is possible to modify the disentangled variables to obtain new generated (\ref{ssec:genration}) and counterfactual images \cite{Cohen2021GifsplanationX-rays, Schutte2021UsingImages}(Section \ref{ssec:counter}).

The computation of the parameters requires optimization through training of the posterior probability, $p_\theta(\mbfz|\mbfx,\mbfy)$, which is intractable. To tackle this issue, we adapt the particular factorization in \equationref{eq:factorization}. We employ deep Variational Autoencoders (deep VAEs) with a bigger expressiveness than traditional VAEs \cite{Kingma2016ImprovingFlow, Child2020VeryImages}. Thus, we can generate more detailed images and implement our hierarchical model.\\
In this way, we obtain the best approximate amortized posterior distribution, $q_\phi(z|x)$, being $\phi$ the parameters of the encoder. Notice that the distribution is amortized just from $\mbfx$ (not from $\mbfy$), so we force the model to extract the meaningful mechanism to generate the segmentation masks just from the self-supervisory signal of the image \cite{LeCun2021Self-supervisedIntelligence}. Indeed, we add a segmentation branch in the architecture (\figureref{fig:arq}), dependent on the main branch.

Namely, we adopt the Noveau VAE (NVAE) \cite{Vahdat2020NVAE:Autoencoder}. This architecture is carefully designed for hierarchical models. Moreover, it has proven efficacy in approximating posteriors by introducing an inductive bias in the image generating process in a deeply hierarchical architecture.

To this aim, the set of $\mbfz$ variables at each representation level $k$ is divided into smaller sets, $m_{k}$, to get a total of $M$ groups of latent variables. Thus, a hierarchical structure is established within each resolution too, being $\mbfz$ the set:
\begin{equation}
\Scale[0.8]{
\mathbf{z} = \big\{ \{(\mathbf{z}_A, \mathbf{z}_S, \mathbf{z}_D)_0,\mathbf{z}_1,\mathbf{z}_2\,...,\mathbf{z}_{m_{k=0}}\}^0, \{(\mathbf{z}_L, \mathbf{z}_K)_{m+1},...,\mathbf{z}_{m_{k=1}}\big\}^1,..., \{ \mathbf{z}_{m+1}, ...,\mathbf{z}_{m_{k}}\}^k,\{\mathbf{z}_{m+1},...,\mathbf{z}_{M}\}^K \big\}
}
\end{equation}
Its prior and approximate posterior probability are given by:
\begin{equation}
    p_\theta(\mbfz) = \prod_m p_\theta(\mathbf{z}_m|\mathbf{z}_{m-1}) \qquad q_\phi(\mbfz|\mbfx) = \prod_m q_\phi(\mathbf{z}_m|\mathbf{z}_{m-1},\mbfx).
\end{equation}
Following this formulation, from marginalization of the $\log$ \equationref{eq:factorization} and rearranging terms, we obtain the variational lower bound to optimize (subscripts colors denote each optimization branch):
\begin{equation} \label{eq:loss}
\small
    \mathcal{L(\mbfx,\mbfy)} = \mathbb{E}_{q_\phi(\textcolor{blue}{\mbfz}|\mbfx)}\big[ log p_\theta(\mbfx|\textcolor{blue}{\mbfz}) \big] -KL(q_\phi(\mathbf{z}_0 | x)||p_\theta(\mathbf{z}_0)) + \mathbb{E}_{q_\phi(\textcolor{magenta}{\mbfz}|\mbfx)}\big[ log p_\theta(\mbfy|\textcolor{magenta}{\mbfz}) \big] - \mathbb{E}_{\textcolor{blue}{\mbfz}} \big[KL_{\textcolor{blue}{\mbfz}} \big] - \mathbb{E}_{\textcolor{magenta}{\mbfz}} \big[KL_{\textcolor{magenta}{\mbfz}} \big] 
\end{equation}
\normalsize
$KL$ being the Kullback–Leibler divergence and
\vspace{-0.5em}
\begin{equation}
\small
    \mathbb{E}_{\mbfz} \big[KL_{\mbfz} \big] = \sum_{m}^{M} \mathbb{E}_{q_\phi(\mathbf{z_{m-1}}|\mbfx)} \big[ KL(q_\phi(\mathbf{z}_{m}| \mathbf{z}_{m-1}, \mbfx) || p_\theta(\mathbf{z}_m|\mathbf{z}_{m-1}) ) \big],
\end{equation}
being $q_\phi(\mathbf{z}_{m-1}|\mbfx)$ the approximate posterior through the hierarchy of $m_{k-1}$ group.\\
Since NVAE convergence depends on the reasonable approximation of KL terms (see \cite{Vahdat2020NVAE:Autoencoder}), to this aim, all priors and posterior probabilities are approximated as Normal distributions. Thus, we can write:
\begin{equation}\label{eq:prior_approx}
    p(\mathbf{z}_A^0) \sim \mathcal{N}(\mu(a), \sigma(a)); \qquad p(\mathbf{z}_S^0) \sim \mathcal{N}(\mu(s), \sigma(s));  \qquad   p(\mathbf{z}_D^0) \sim \mathcal{N}(\mu(d), \sigma(d));
\end{equation}

\section{Experiments and Results}

\subsection{Datasets description}\label{ssec:datasets}

The model is optimized employing small datasets: ten lung CT volumes per animal model ($\sim2000$ slices). The data used for training are axial slices from three \Mtb~lung models identified as follows.

The dataset names identify: the animal model, $A$, the data source and the phase as follows $A_{phase}^{source}$). Namely, the human volumes, $H_{tr}^{CLE}$, corresponds to the validation data of the 2019 ImageClefMed TB task \cite{DicenteCid2019OverviewAssessment}. The mice images, $M_{tr}^{GSK}$, are provided from \textit{GlaxoSmithKline plc.} (GSK) within the context of the ERA4TB project \cite{ERA4TBconsotium2021ERA4TB}, similarly to the primate ones, $P_{tr}^{PHE}$, from the \textit{Public Health of England} (PHE) \cite{Gordaliza2018, Gordaliza2019AManifestations}.
For testing (twenty volumes per model), $P_{ts}^{PHE}$ and $M_{ts}^{GSK}$, are selected from different cohorts of $P_{tr}^{PHE}$ and $P_{tr}^{GSK}$, while the human dataset, $H_{ts}^{CLE}$ is a partition of the mentioned data. The remaining sets are included to evaluate the model generalisation and transferability capabilities. $M_{ts}^{EXM}$ belongs to a public dataset from the Institute for Experimental Molecular Imaging (ExMI) \cite{Rosenhain2018ASegmentations} which contains healthy subjects at low resolution. Finally, the human dataset, $H_{ts}^{RAD}$, presents subjects with lung damage caused by COVID-19 \cite{Cohen2020COVID-19Collection}.

Note that all datasets include segmentation masks delineated by trained experts.

A detailed description of the different datasets is presented in \tableref{tbl:datasets}.

\begin{table}[htpb]
\centering
\caption{Datasets description }
\label{tbl:datasets}
\arrayrulecolor{white}
\scalebox{0.6}{
\begin{tabular}{c||c||c||c||c||c||c}
\rowcolor[rgb]{0.2,0.2,0.6} \textcolor{white}{\textbf{Dataset ID}} & \textcolor{white}{\textbf{Phase}} & \textcolor{white}{\textbf{Animal Model}} & \textcolor{white}{\textbf{Source}} & \textcolor{white}{\# Slices} & \textcolor{white}{\textbf{Voxel Spacing [mm]}} & \textcolor{white}{\textbf{Resolution~}} \\ 
\hhline{=::=::=::=::=::=::=}
\rowcolor[rgb]{0.945,0.945,0.945} $M^{GSK}_{tr}$ & \textcolor[rgb]{0.259,0.259,0.259}{Training} & {\cellcolor[rgb]{0.945,0.945,0.945}} & {\cellcolor[rgb]{0.945,0.945,0.945}} & \textcolor[rgb]{0.259,0.259,0.259}{2002} & {\cellcolor[rgb]{0.945,0.945,0.945}} & {\cellcolor[rgb]{0.945,0.945,0.945}} \\ 
\hhline{-||-||>{\arrayrulecolor[rgb]{0.945,0.945,0.945}}->{\arrayrulecolor{white}}||>{\arrayrulecolor[rgb]{0.945,0.945,0.945}}->{\arrayrulecolor{white}}||-||>{\arrayrulecolor[rgb]{0.945,0.945,0.945}}->{\arrayrulecolor{white}}||>{\arrayrulecolor[rgb]{0.945,0.945,0.945}}-}
\rowcolor[rgb]{0.945,0.945,0.945} $M^{GSK}_{ts}$ & {\cellcolor[rgb]{0.945,0.945,0.945}} & {\cellcolor[rgb]{0.945,0.945,0.945}} & \multirow{-2}{*}{{\cellcolor[rgb]{0.945,0.945,0.945}}\textcolor[rgb]{0.259,0.259,0.259}{GSK }} & \textcolor[rgb]{0.259,0.259,0.259}{3987} & \multirow{-2}{*}{{\cellcolor[rgb]{0.945,0.945,0.945}}$0.087 \times 0.087$} & \multirow{-2}{*}{{\cellcolor[rgb]{0.945,0.945,0.945}}$500 \times 500$} \\ 
\hhline{>{\arrayrulecolor{white}}-||>{\arrayrulecolor[rgb]{0.945,0.945,0.945}}->{\arrayrulecolor{white}}||>{\arrayrulecolor[rgb]{0.945,0.945,0.945}}->{\arrayrulecolor{white}}||-||-||-||-}
\rowcolor[rgb]{0.945,0.945,0.945} $M^{EXM}_{ts}$ & \multirow{-2}{*}{{\cellcolor[rgb]{0.945,0.945,0.945}}\textcolor[rgb]{0.259,0.259,0.259}{Test }} & \multirow{-3}{*}{{\cellcolor[rgb]{0.945,0.945,0.945}}\textcolor[rgb]{0.259,0.259,0.259}{Mouse }} & \textcolor[rgb]{0.259,0.259,0.259}{ExMI} & \textcolor[rgb]{0.259,0.259,0.259}{3785} & $0.282 \times 0.282$ & $144 \times 100$ \\ 
\hhline{=::=::=::=::=::=::=}
\rowcolor[rgb]{0.867,0.867,0.867} $P^{PHE}_{tr}$ & \textcolor[rgb]{0.259,0.259,0.259}{Training} & {\cellcolor[rgb]{0.867,0.867,0.867}} & {\cellcolor[rgb]{0.867,0.867,0.867}} & \textcolor[rgb]{0.259,0.259,0.259}{2012} & {\cellcolor[rgb]{0.867,0.867,0.867}} & {\cellcolor[rgb]{0.867,0.867,0.867}} \\ 
\hhline{-||-||>{\arrayrulecolor[rgb]{0.867,0.867,0.867}}->{\arrayrulecolor{white}}||>{\arrayrulecolor[rgb]{0.867,0.867,0.867}}->{\arrayrulecolor{white}}||-||>{\arrayrulecolor[rgb]{0.867,0.867,0.867}}->{\arrayrulecolor{white}}||>{\arrayrulecolor[rgb]{0.867,0.867,0.867}}-}
\rowcolor[rgb]{0.867,0.867,0.867} $P^{PHE}_{ts}$ & \textcolor[rgb]{0.259,0.259,0.259}{Test} & \multirow{-2}{*}{{\cellcolor[rgb]{0.867,0.867,0.867}}\textcolor[rgb]{0.259,0.259,0.259}{Primate }} & \multirow{-2}{*}{{\cellcolor[rgb]{0.867,0.867,0.867}}\textcolor[rgb]{0.259,0.259,0.259}{PHE }} & \textcolor[rgb]{0.259,0.259,0.259}{4021} & \multirow{-2}{*}{{\cellcolor[rgb]{0.867,0.867,0.867}}$0.235 \times 0.235$} & \multirow{-2}{*}{{\cellcolor[rgb]{0.867,0.867,0.867}}$512 \times 512$} \\ 
\hhline{>{\arrayrulecolor{white}}=::=::=::=::=::=::=}
\rowcolor[rgb]{0.945,0.945,0.945} \textbf{$H^{CLE}_{tr}$~} & \multicolumn{1}{l||}{\textcolor[rgb]{0.259,0.259,0.259}{Training}} & {\cellcolor[rgb]{0.945,0.945,0.945}} & {\cellcolor[rgb]{0.945,0.945,0.945}} & \textcolor[rgb]{0.259,0.259,0.259}{1617} & {\cellcolor[rgb]{0.945,0.945,0.945}} & {\cellcolor[rgb]{0.945,0.945,0.945}} \\ 
\hhline{-||-||>{\arrayrulecolor[rgb]{0.945,0.945,0.945}}->{\arrayrulecolor{white}}||>{\arrayrulecolor[rgb]{0.945,0.945,0.945}}->{\arrayrulecolor{white}}||-||>{\arrayrulecolor[rgb]{0.945,0.945,0.945}}->{\arrayrulecolor{white}}||>{\arrayrulecolor[rgb]{0.945,0.945,0.945}}-}
\rowcolor[rgb]{0.945,0.945,0.945} $H^{CLE}_{ts}$ & {\cellcolor[rgb]{0.945,0.945,0.945}} & {\cellcolor[rgb]{0.945,0.945,0.945}} & \multirow{-2}{*}{{\cellcolor[rgb]{0.945,0.945,0.945}}\textcolor[rgb]{0.259,0.259,0.259}{ImageClef }} & \textcolor[rgb]{0.259,0.259,0.259}{3578} & \multirow{-2}{*}{{\cellcolor[rgb]{0.945,0.945,0.945}}$0.60\text{-}0.75\times0.60\text{-}0.75$} & \multirow{-2}{*}{{\cellcolor[rgb]{0.945,0.945,0.945}}$512 \times 512$} \\ 
\hhline{>{\arrayrulecolor{white}}-||>{\arrayrulecolor[rgb]{0.945,0.945,0.945}}->{\arrayrulecolor{white}}||>{\arrayrulecolor[rgb]{0.945,0.945,0.945}}->{\arrayrulecolor{white}}||-||-||-||-}
\rowcolor[rgb]{0.945,0.945,0.945} $H^{RAD}_{ts}$ & \multirow{-2}{*}{{\cellcolor[rgb]{0.945,0.945,0.945}}\textcolor[rgb]{0.259,0.259,0.259}{Test }} & \multirow{-3}{*}{{\cellcolor[rgb]{0.945,0.945,0.945}}\textcolor[rgb]{0.259,0.259,0.259}{Human }} & \textcolor[rgb]{0.259,0.259,0.259}{Radiopedia} & \textcolor[rgb]{0.259,0.259,0.259}{4034} & $0.68\text{-}0.75\times0.68\text{-}0.75$ & $512\text{-}630\times430\text{-}630$
\end{tabular}
}
\arrayrulecolor{black}
\end{table}
\subsection{Implementation details} \label{ssec_implemtation}
The model is optimized employing six scales, $K=6$, with $18$ latent variables per scale, partitioned each in $m_k$ groups as follows, $m_k = [2,2,2,3,6,9]$. 
The three $\mu_A$, $\mu_S$ and $\mu_D$ are known during training ($\mu_A=[-1,0,1]$, $\mu_D=(0,1)$, $\mu_S=(0,1)$), fix at image generation and inferred for image reconstruction and segmentation mask generation employing $KL\big(q_\phi(z^0) || \mathcal{N}(0,1)\big)$. During optimization $\mu_D$ is given by the the healthy lung relative volume (extracted by simple thresholding) with respect to the ground truth mask volume.

\subsection{Pathological Lungs Generation}
\label{ssec:genration}

After optimization, 
the model can generate realistic images, such as those shown in \figureref{fig:generated}, by choosing the mean values of $\mathbf{z}_A^0$, $\mathbf{z}_S^0$, $\mathbf{z}_D^0$ factors. To illustrate this capacity in \figureref{fig:generated}, we set a relative slice position of $0.5$, the animal model is fixed for each row and, the effect of the lung damage variable is modulated from lower to higher in each column. \\

\begin{figure}[htbp]
\floatconts
  {fig:generated}
  {\caption{\small Synthetic lung CT images generated by our model. Images are generated with a fixed slice relative position ($\mu_S$). For each row, the animal model $\mu_A$ is fixed to $-1,0,1$, respectively, while for each column, the damage $\mu_D$ is increased [$0$-$1$]. }}
  {\includegraphics[width=0.9\linewidth, height=0.3\textheight]{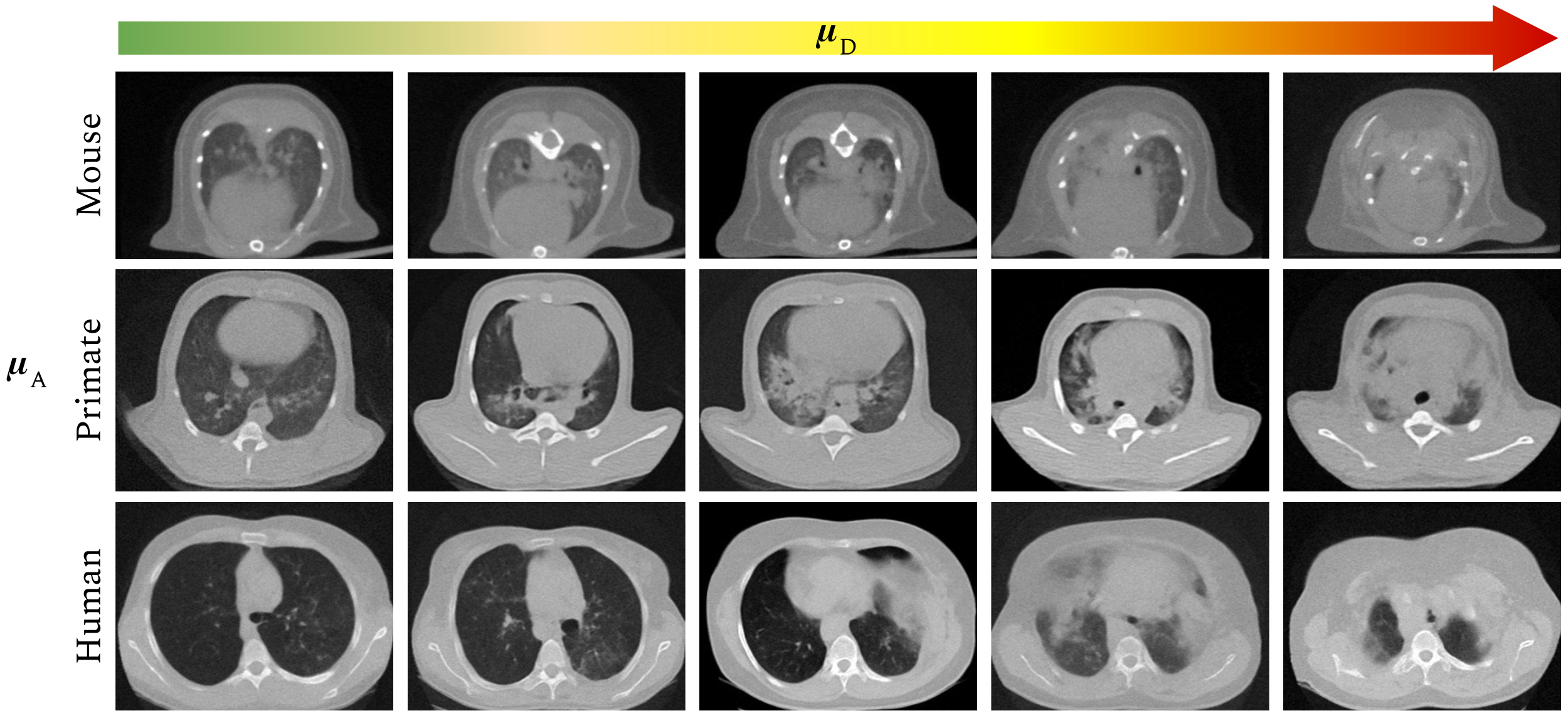}}
\end{figure}

\vspace{-2em}
\subsection{Counterfactual Images}\label{ssec:counter}

The first column of each row in \figureref{fig:counter} shows an actual image of a damaged lung corresponding to a given animal model. When no actions are performed, the model infers the disentangled image representation of the causative variables ($\mathbf{z}_A^0$, $\mathbf{z}_S^0$, $\mathbf{z}_D^0$) through the encoder. Subsequently, the image is reconstructed, and a segmentation mask (third column) is generated employing the optimized decoder (\figureref{fig:arq}).
The second column shows a healthy counterfactual of the input images, which is generated setting to zero the mean value of the inferred damage variable, $\mathbf{z_D^0}$. The decoder is fed with the zero-mean  $\mathbf{z_D^0}$ and the rest (unaltered) inferred causal variables to generate the counterfactual version of the slice and its respective mask (fourth column).

\begin{figure}[htbp]
\floatconts
  {fig:counter}
  {\caption{The encoder infers the real image (axial slice) disentangled representation, $\mathbf{z}_A^0$, $\mathbf{z}_S^0$, $\mathbf{z}_D^0$. By setting the damage variable $\mathbf{z}_D^0$ to $0$ the decoder generates the healthy counterfactual (counterfactual slice) and its respective mask (counterfactual mask).} }
  {\includegraphics[width=0.9\linewidth, height=0.3\textheight]{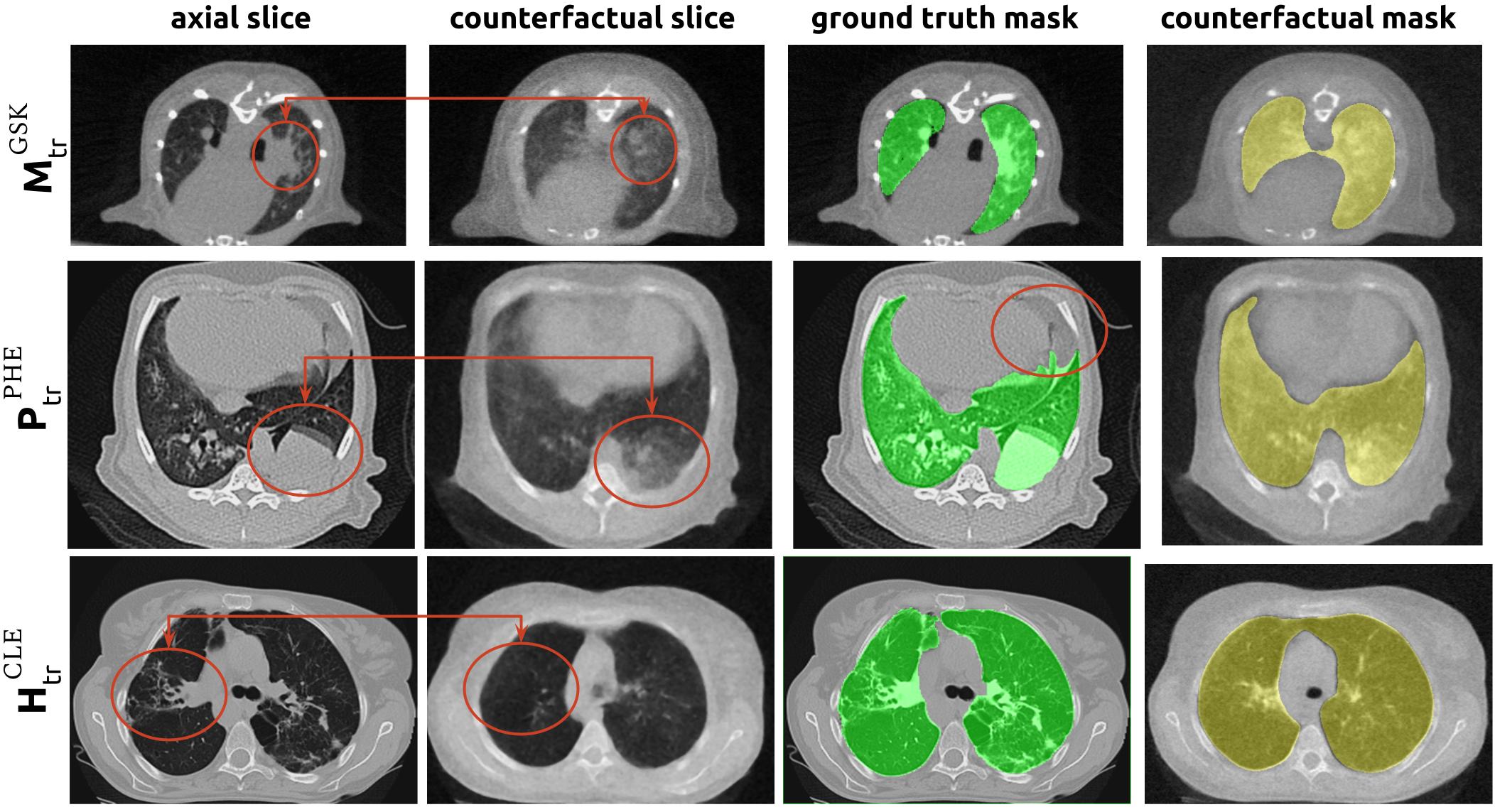}}
\end{figure}
\vspace{-1em}

\subsection{Segmentation employing counterfactual images}

Pathological lung segmentation is an important task to solve in drug development studies. Unfortunately, it is a complex task due to the difficulty of discrimination between lesions and other neighborhood tissues. Needless to say that the diversity of the biological data supposes an added difficulty\cite{Hofmanninger2020AutomaticProblemb}. In this experiment, we retrain the optimized model with counterfactual images to generate the segmentation masks from the test datasets (Section \ref{ssec:datasets}). We use the approach described previously to generate the counterfactual images (Section \ref{ssec:counter}). 
To learn about the strengths and weaknesses of this generative approach, we compare the results obtained, $our_{c}$, with the segmentation masks calculated by our original method, $our_{nc}$, and the state-of-the-art fully supervised method, \textit{nnU-Nnet} \cite{Isensee2021NnU-Net:Segmentation}.

\begin{table}[htpb]
\centering
\caption{\small Mean and standard deviation (SD) of the Dice Similarity Coefficient (DSC) and Hausdorff Distance (HD) between the ground truth masks and mask obtained from the methods indicated at rows (\textit{nnU-Nnet}, proposed method before employing counterfactual images ($our_{nc}$), and after ($our_{c}$)) for each test dataset (columns).}
\label{tbl:optimiza_results}
\arrayrulecolor{white}
\scalebox{0.55}{
\begin{tabular}{c||c|c|c|c|c||c|c|c|c|c}
\multicolumn{1}{c}{} & \multicolumn{5}{c||}{{\cellcolor[rgb]{0.322,0.322,0.322}}\textcolor{white}{DSC $\pm$ SD}} & \multicolumn{5}{c}{{\cellcolor[rgb]{0.322,0.322,0.322}}\textcolor{white}{HD $\pm$ SD [mm]}} \\ 
\hhline{~-----||-----}
\multicolumn{1}{c}{} & {\cellcolor[rgb]{0.6,0.6,0.6}}$M^{GSK}_{ts}$ & {\cellcolor[rgb]{0.6,0.6,0.6}}$M^{EXT}_{ts}$ & {\cellcolor[rgb]{0.6,0.6,0.6}}$P^{PHE}_{ts}$ & {\cellcolor[rgb]{0.6,0.6,0.6}}$H^{CLE}_{ts}$ & {\cellcolor[rgb]{0.6,0.6,0.6}}$H^{COV}_{ts}$ & {\cellcolor[rgb]{0.6,0.6,0.6}}$M^{GSK}_{ts}$ & {\cellcolor[rgb]{0.6,0.6,0.6}}$M^{EXM}_{ts}$ & {\cellcolor[rgb]{0.6,0.6,0.6}}$P^{PHE}_{ts}$ & {\cellcolor[rgb]{0.6,0.6,0.6}}$H^{CLE}_{ts}$ & {\cellcolor[rgb]{0.6,0.6,0.6}}$H^{COV}_{ts}$ \\ 
\hhline{>{\arrayrulecolor[rgb]{0.973,0.973,0.973}}->{\arrayrulecolor{white}}|t|>{\arrayrulecolor[rgb]{0.973,0.973,0.973}}---->{\arrayrulecolor{white}}-||>{\arrayrulecolor[rgb]{0.973,0.973,0.973}}-----}
\rowcolor[rgb]{0.973,0.973,0.973} \textcolor[rgb]{0.2,0.2,0.2}{nnU-Net } & \textcolor[rgb]{0.2,0.2,0.2}{}$0.845 \pm 0.10$\textcolor[rgb]{0.2,0.2,0.2}{} & \textcolor[rgb]{0.2,0.2,0.2}{$0.851 \pm 0.11$} & \textcolor[rgb]{0.2,0.2,0.2}{$0.957 \pm 0.06$} & \textcolor[rgb]{0.2,0.2,0.2}{$0.978 \pm 0.04$} & \textcolor[rgb]{0.2,0.2,0.2}{$0.973 \pm 0.03$} & \textcolor[rgb]{0.2,0.2,0.2}{$1.737 \pm 1.01$} & \textcolor[rgb]{0.2,0.2,0.2}{$1.90 \pm 1.52$} & \textcolor[rgb]{0.2,0.2,0.2}{$3.30 \pm 3.96$} & $9.37 \pm 15.14$ & \textcolor[rgb]{0.2,0.2,0.2}{$8.31 \pm 10.71$} \\ 
\arrayrulecolor{white}\hline
\rowcolor[rgb]{0.867,0.867,0.867} \textcolor[rgb]{0.2,0.2,0.2}{$our_{nc}$} & $0.849 \pm 0.10$ & \textcolor[rgb]{0.2,0.2,0.2}{$0.843 \pm 0.12$} & \textcolor[rgb]{0.2,0.2,0.2}{$0.949 \pm 0.06$} & \textcolor[rgb]{0.2,0.2,0.2}{$0.963 \pm 0.06$} & \textcolor[rgb]{0.2,0.2,0.2}{$0.963 \pm 0.06$} & \textcolor[rgb]{0.2,0.2,0.2}{$1.948 \pm 1.11$} & \textcolor[rgb]{0.2,0.2,0.2}{$2.06 \pm 1.82$} & \textcolor[rgb]{0.2,0.2,0.2}{$3.81 \pm 4.10$} & \textcolor[rgb]{0.2,0.2,0.2}{$10.12 \pm 18.32$} & \textcolor[rgb]{0.2,0.2,0.2}{$10.56 \pm 10.77$} \\ 
\hline
\rowcolor[rgb]{0.973,0.973,0.973} \textcolor[rgb]{0.2,0.2,0.2}{$our_{c}$ } & \textcolor[rgb]{0.2,0.2,0.2}{$0.877 \pm 0.08$} & \textcolor[rgb]{0.2,0.2,0.2}{$0.859 \pm 0.11$} & \textcolor[rgb]{0.2,0.2,0.2}{$0.955 \pm 0.06$} & \textcolor[rgb]{0.2,0.2,0.2}{$0.977 \pm 0.06$} & \textcolor[rgb]{0.2,0.2,0.2}{$0.968 \pm 0.04$} & \textcolor[rgb]{0.2,0.2,0.2}{$1.519 \pm 0.89$} & \textcolor[rgb]{0.2,0.2,0.2}{$1.88 \pm 1.53$} & \textcolor[rgb]{0.2,0.2,0.2}{$2.95 \pm 3.54$} & \textcolor[rgb]{0.2,0.2,0.2}{$8.78 \pm 16.11$} & \textcolor[rgb]{0.2,0.2,0.2}{$9.48 \pm 9.89$}
\end{tabular}
}
\arrayrulecolor{black}
\end{table}

\tableref{tbl:optimiza_results} shows the mean and standard deviation for Dice Similarity Coefficient (DSC) and Hausdorff Distance (HD) between each segmentation method and the ground truth masks for each test dataset. The results present an improvement for all measures and datasets when employing counterfactual images, yielding similar results to the \textit{nnU-Nnet}. The differences are due to subtle changes in most of the cases or even small imperfections in the ground truth masks as it is shown in \figureref{fig:lungs_seg}.  
\begin{figure}[htbp]
 
\floatconts
  {fig:lungs_seg}
  {\caption{\small Comparison of methods. Each row contains axial slices and segmentation masks of each test dataset. Columns show the original image, ground truth mask (green), \textit{nnU-Net} mask (blue), overlay of \textit{nnU-Net} and ground truth (cyan), the mask with our method employing counterfactual images during training (yellow) and the overlay with the ground truth (lime). Red and green circles show inaccuracies and precise segmentation cases, respectively.}}
  {\includegraphics[width=0.85\linewidth, height=0.4\textheight]{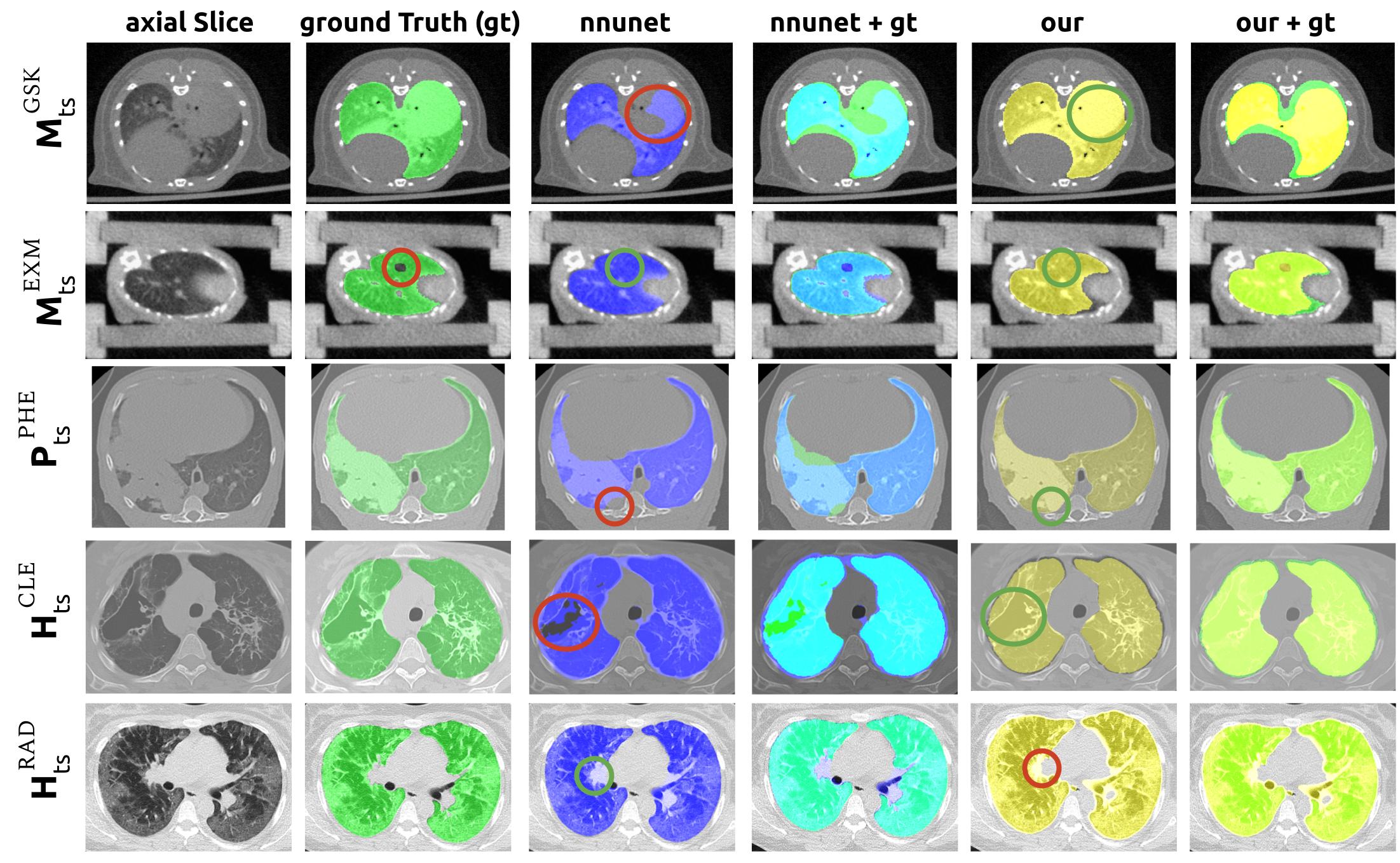}}
\end{figure}

\vspace{-2em}
\section{Conclusions}
The methodology proposed in this work yields promising results obtaining the factors characterizing the pathophysiological processes shared between animal models. 
Although the approach indeed suffers from several limitations: the use of isolated axial slices instead of the more informative whole three-dimensional images and the characterization of disease affectation based simply on the damaged lung volume and not on the specific manifestations of the disease for each animal model. These limitations will be the object of future work.

To sum up, our model is capable of inferring meaningful disentangled representations.
Namely, it generates synthetic slices by setting the values of the modelled factors. Even more relevant, it produces counterfactual versions of existing slices by testing the effective disentanglement. In the future, we explore strategies to exploit the approach to increase the diversity of existing data, essential for automatic segmentation, or to provide the damage variable as a possible (to be validated) inter-species biomarker.

\midlacknowledgments{This project has received funding from the Innovative Medicines Initiative 2 Joint Undertaking (JU) under grant agreement No. 853989. The JU receives support from the European Union’s Horizon 2020 research and innovation programme and EFPIA and Global Alliance for TB Drug Development non-profit organisation, Bill \& Melinda Gates Foundation and University of Dundee. This work was partially funded by Ministerio de Ciencia, Innovación y Universidades, Agencia Estatal de Investigación, under grant PID2019-109820RB-I00, MCIN/AEI/10.13039/501100011033/, co-finance by European Regional Development Fund (ERDF), “A way of making Europe.”}

\bibliography{references}

\clearpage
\appendix

\clearpage

\section{Life cycle of Tuberculosis infection}\label{ap:cycle}

\begin{figure}[htpb]
\centering
\includegraphics[width=0.75\textwidth,height=0.4\textheight]{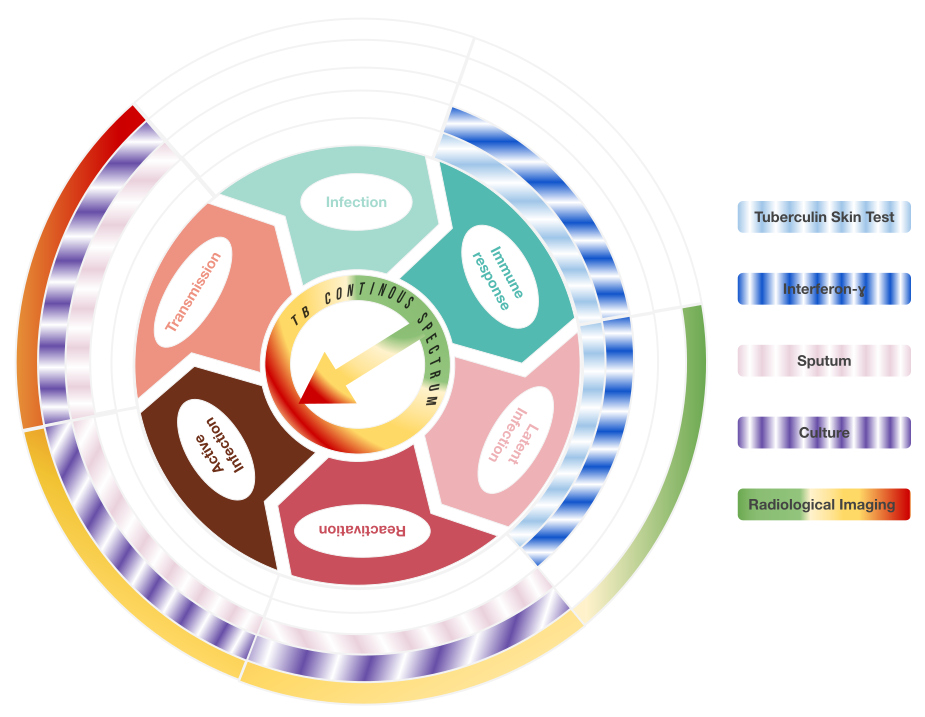}
\caption[Life cycle of \Mtb and main tests]{Life cycle of \Mtb~\cite{Arroyo-Ornelas2012ImmuneTechnologies,Ernst2012TheTuberculosis} and main tests to characterise the entire disease spectrum. The inner cycle names the traditional categorical clinical stages of the continuous spectrum of TB immunological life cycle. Each outer circle represent each TB assessment tests capability. Blank spaces for lack of sensibility, bicolour ones represent the binary character of the test, while gradient representation represents the ability to provide a continuous value.} 
\end{figure}\label{fig:TB_tests}

\section{Extra Experiments Setup details}

The following list offers further details about the context of the experiments.
\begin{itemize}
    \item \textbf{System setup:} All experiments were performed in a machine with an Intel Xeon 8153 CPU, 64-GB RAM and two 12-GB Titan V GPUs. We created a specific Docker image based on Ubuntu 20.04 with Python 3.6.9 and torch 1.6.0 to run our code.  
    
    \item \textbf{Preprocessing:} To reduce the size of the chest CT images, we crop the images and their respective segmentation masks to the body region. We employ thresholding from $-1024$ to $600$ over the Hounsfield Units (HU) followed by morphological operations to eliminate small isolated blobs. Finally, we select the one corresponding to the whole body region. 
    
Since \textit{nnU-Net} automatically estimates the rest of preprocessing operations, these cropped volumes feed the \textit{nnU-Net} preprocessing pipelines. The details about \textit{nnU-Net} experiments are given below in the list.

In the case of our model, we rescale the cropped images resolution to 256 x 256 pixels and normalize the intensity (0-1).

During training, our model needs an estimation of healthy lung volume per CT (Sections \ref{sec:Methods} and \ref{ssec_implemtation}). To this aim, over the cropped image, we apply a threshold to recover just the healthy tissue inside the whole lung mask. Following the experts' recommendations, we set this threshold from $-900$ to $-200$ HUs for the human training dataset ($H^{CLE}_{tr}$), -1000 to -200 HUs for the macaque dataset ($P^{PHE}_{tr}$), and from -800 to -300 HUs for the mouse model ($M^{GSK}_{tr}$). The healthy volume extracted is divided by the total mask volume to obtain the relative value employed during training.   
    
    \item \textbf{Selection of each dataset sample:} We use $30$ CT volumes per dataset employed during training and testing and $20$ when the datasets are employed just during the test phase, as described in Section \ref{ssec:datasets}.
    
    Except for the $M^{GSK}$ and $H^{RAD}$ datasets, the rest of the original datasets contain more than $30/20$ volumes. 
    
To define our specific trimmed samples, we employ the relative healthy volume to classify each CT as low damage (relative healthy volume $\geq 0.85$), medium damage ($0.85 >$ relative healthy volume $> 0.4$) and high damage (relative healthy volume $\leq 0.4$). Subsequently, we randomly select the same number of subjects per interval. 
    
    \item \textbf{Training details:} We employ the two Titan V GPUs during $900$ epochs with a total batch size of $8$ using the Adamax optimizer with an initial learning rate of $0.01$ and \textit{Cosine Annealing} scheduler (minimal learning rate: $1e-4$). We apply online data augmentation to the normalized images by employing random affine transformations (10º rotation) and adding Gaussian noise ($\mu=0$, $\sigma=0.05$).  
    
    For the \textit{nnU-Net} \cite{Isensee2021NnU-Net:Segmentation}, we use a single Titan V GPU, following the nnU-Net authors'  (\href{https://github.com/MIC-DKFZ/nnUNet}{recommendations}). After adapting the cropped image name formats to the nnU-Net requirements, we run the \texttt{nnUNet\_plan\_and\_preprocess} function to allow the pipeline to estimate the network configuration and training parameters. The complete list can be found in the following \href{https://drive.google.com/file/d/1-fmQ_tt1ElHhO6rHMzsUG7QYLaOszn5C/view?usp=sharing}{link}. Subsequently, we train a 2D configuration in a 5-fold cross-validation during $1000$ epochs per fold, employing a batch size of $14$, data augmenting (see \href{https://drive.google.com/file/d/1-fmQ_tt1ElHhO6rHMzsUG7QYLaOszn5C/view?usp=sharing}{linked} file for details), the SGD optimizer and a learning rate of $0.01$
    
    \item \textbf{Image Generation speed:} After loading the trained model ($\sim 20s$), it is possible to generate a 16 batch size of $256 \times 256$ images in approximately 0.25s.

\end{itemize}

\clearpage
\section{Pathological Lungs Generation: Varying the slice position}
This appendix shows generated slices instances fixing the damage and varying the relative slice position. This experiment extends Section \ref{ssec:genration}, in which axial slices belong to a fixed relative slice position.

Since our chest CT volumes orientation is cephalic to caudal, the model generates axial images of the upper airways (trachea) and the corresponding per animal model surrounding tissues at the lowest slice position, as shown in the first column of the \figureref{fig:generated_vary}. This way, the second column shows the corresponding generated anatomy for the superior lungs, while the third and fourth columns accordingly show the middle and inferior regions. Finally, the fifth column depicts the generated version at the beginning of the abdominal anatomy. 

\begin{figure}[htbp]
\floatconts
  {fig:generated_vary}
  {\caption{Synthetic lung CT images generated by our model. Images are generated with a fix relative damage, $\mu_D=0.5$. For each row, the animal model $\mu_A$ is fixed to $-1,0,1$, respectively, while for each column, the relative slice position $\mu_S$ is increased between $0$ and $1$. }}
  {\includegraphics[width=0.9\linewidth, height=0.3\textheight]{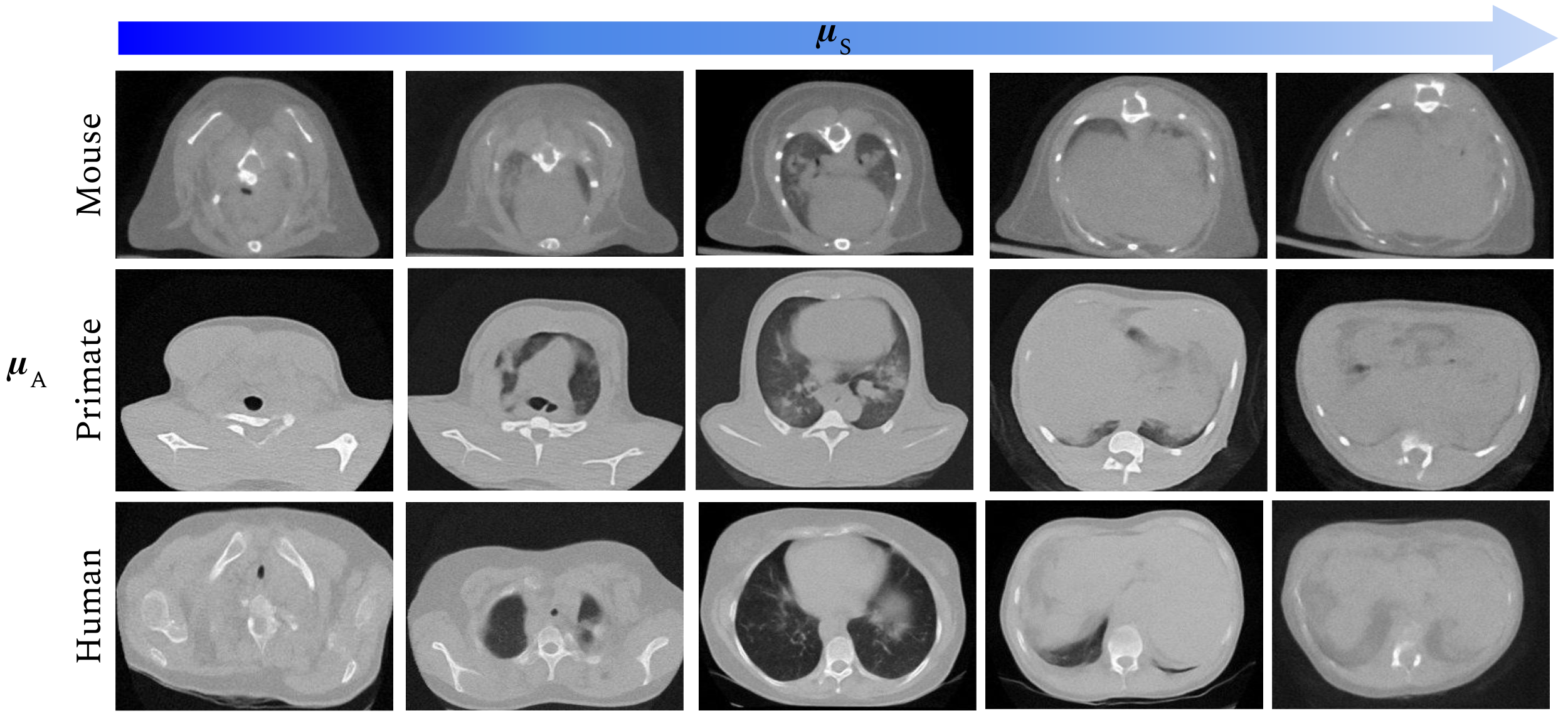}}
\end{figure}\label{fig:var_slic}

\clearpage
\section{Counterfactual Images: Extended Assessment}
This appendix extends the qualitative results presented in Section \ref{ssec:counter}. The former section shows the model capacity generating counterfactual images and their respective segmentation masks.

Here, we evaluate how realistic are the generated images. For that, we compare the Hounsfield Units (HU) of real CT slices with two cases: a) the reconstructed slice from the variable inferred by the encoder without modification of any of these values, and b) the counterfactual image, namely, after intervening on the inferred damage value. 
We compute the voxel-wise Root Mean Square Error (RMSE) for the reconstructed images per test dataset. \tableref{tbl:RMSE} shows these results with an average $RMSE=18.73\pm2.16$. 

Voxel-wise evaluation is not suitable for counterfactual images. Previous manual delimitation of comparable regions is needed, which is a priority for our future work.

To illustrate similarities and differences in the HU scale, in \figureref{fig:profiles}, we plot the HU profile belonging to the damaged regions shown in \figureref{fig:counter}. Respectively, the first three rows contain 1) the original axial slice from the different test datasets (the image is generated from the $\mu_a$, $\mu_s$ and $\mu_d$ inferred by our model), with the profile horizontal line in green, 2) the reconstructed slice (the image is generated maintaining $\mu_a$, $\mu_s$ inferred by our model and correcting $\mu_d$), with profile line in yellow and 3) the counterfactual after modifying the inferred expected damage, with the profile line in blue. 

The last row shows the HU plot for each profile-specific colour.   
HU values are similar for the three slices except for those regions where the slice counterfactual version replaces the damage with healthy tissue-like. We highlight such changes framing them in vertical dashed red lines.

Besides, it is important to note that the original and reconstructed images present more noisy patterns than the counterfactual version, as was expected from its blurrier appearance and the thickening of the soft tissue for the mice dataset. 

\begin{table}[htpb]
\centering
\caption{Root Mean Square Error (RMSE) between the real images and the image reconstructed from the $\mu_a$, $\mu_s$ and $\mu_d$ inferred by our model for the test datasets}
\label{tbl:RMSE}
\arrayrulecolor{white}
\begin{tabular}{c|c||c||c|c|}
\multicolumn{5}{c}{{\cellcolor[rgb]{0.322,0.322,0.322}}\begin{tabular}[c]{@{}>{\cellcolor[rgb]{0.322,0.322,0.322}}c@{}}\textcolor{white}{RMSE}\textcolor{white}{ [HU]}\\\end{tabular}} \\ 
\hhline{==:t:=:t:==|}
\rowcolor[rgb]{0.6,0.6,0.6} $M^{GSK}_{ts}$ & $M^{EXT}_{ts}$ & $P^{PHE}_{ts}$ & $H^{CLE}_{ts}$ & $H^{COV}_{ts}$ \\ 
\hline
$21.26$ & {\cellcolor[rgb]{0.973,0.973,0.973}}\textcolor[rgb]{0.2,0.2,0.2}{$18.75$} & {\cellcolor[rgb]{0.973,0.973,0.973}}\textcolor[rgb]{0.2,0.2,0.2}{$20.12$} & {\cellcolor[rgb]{0.973,0.973,0.973}}\textcolor[rgb]{0.2,0.2,0.2}{$17.89$} & {\cellcolor[rgb]{0.973,0.973,0.973}}\textcolor[rgb]{0.2,0.2,0.2}{$15.63$} \\
\hhline{~-|b|-|b|--|}
\end{tabular}
\arrayrulecolor{black}
\end{table}

\begin{figure}[htbp]
\floatconts
  {fig:profiles}
  {\caption{Hounsfield Units (HU) plots for profiles at regions damaged in original test axial slices. Each column contains instances of each dataset, previously employed in Section \ref{ssec:counter}. The first rows depict the original, reconstructed and counterfactual slices with the profile line green, yellow and blue, respectively. The last row draws the HU profiles per voxel. Vertical dashed lines highlight big differences between real/reconstructed and counterfactual slices.}}
  {\includegraphics[width=\linewidth, height=0.55\textheight]{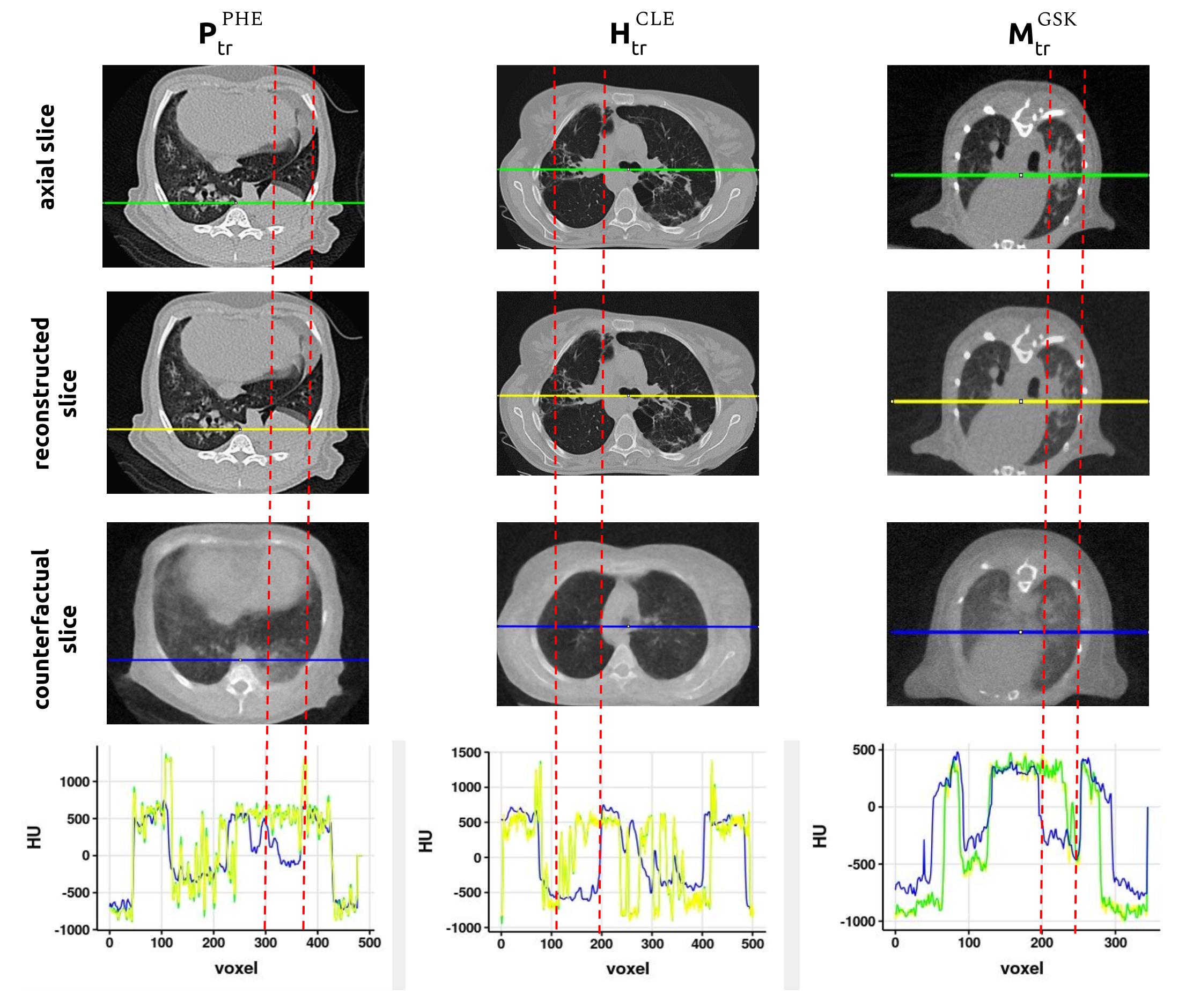}}
\end{figure}

\end{document}